\documentclass[prd,superscriptaddress,showpacs,twocolumn,amsmath,amssymb]{revtex4}
\usepackage{graphicx}
\usepackage{amsmath}
\usepackage{amssymb}
\usepackage{float}
\usepackage{color}

\usepackage{cancel,xcolor}
\usepackage{indentfirst}
\usepackage{CJKulem}
\usepackage{soul,xcolor}
\setstcolor{red}
\usepackage{ulem}
\usepackage{cancel}
\usepackage[pdfpagemode=UseNone,pdfstartview=FitH,colorlinks=true,linkcolor=blue,urlcolor=blue,anchorcolor=blue,citecolor=blue]{hyperref}

\newcommand\bea{\begin{eqnarray}}
\newcommand\eea{\end{eqnarray}}
\newcommand\beq{\begin{equation}}  
\newcommand\eeq{\end{equation}}

\begin{document}

\title{\titlename}
\title{Timelike entanglement and central charge for quantum BTZ black holes}
\author{Dibakar Roychowdhury}
\email{dibakar.roychowdhury@ph.iitr.ac.in}
\affiliation{Department of Physics, Indian Institute of Technology Roorkee
Roorkee 247667, Uttarakhand,
India}

\begin{abstract}
We compute holographic timelike entanglement entropy for quantum BTZ black holes in a Karch-Randall braneworld scenario. These black holes are exact solutions of massive 3d gravity on the brane and are conjectured to be dual to thermal states in 2d defect CFTs, living at the interface of the brane and the boundary of the bulk $AdS_4$. Our analysis reveals an interesting relation between the tEE and the central charge pertaining to the dual CFT, which receives nontrivial corrections due to quantum backreaction effects on the Karch-Randall brane, which is explored non-perturbatively.
\end{abstract}
\maketitle
\section{Introduction and motivation}
Quantum black hole solutions in three dimensions \cite{Emparan:2002px}-\cite{Panella:2023lsi} and their charged \cite{Climent:2024nuj}-\cite{Feng:2024uia} and rotating black hole \cite{Emparan:2020znc}-\cite{Bhattacharya:2025tdn} generalizations have garnered renewed interest in recent years due to various exciting features \footnote{For a nice and comprehensive review, see \cite{Panella:2024sor}. The reader is also referred to \cite{Cartwright:2024iwc}-\cite{Chen:2023tpi} for various other interesting developments that took place in recent years.}. It has been a long-standing problem to address the issue of black hole formation in the presence of large quantum backreaction effects \cite{Steif:1993zv}-\cite{Casals:2019jfo}, which becomes even more challenging in the presence of large number of fields where the perturbative techniques break down. Braneworld  holography \cite{deHaro:2000wj} provides a natural platform to address such issues, where the backreaction problem can be solved exactly. In braneworld holography, the IR cut-off surface in the bulk is replaced by an end of the world (ETW) brane, namely, the Randall-Sundrum (RS) \cite{Randall:1999ee}-\cite{Randall:1999vf} or Karch-Randall (KR) \cite{Karch:2000ct}-\cite{Karch:2000gx} brane placed at a finite distance from the boundary at asymptotic infinity. 

Given the bulk as $AdS_4$, we have a new massive $AdS_3$ gravity on the KR brane, fully backreacted by the quantum conformal matter (which is the 3d boundary CFT from the perspective of the bulk $AdS_4$) living on the brane. The graviton on the brane receives mass due to quantum bacreaction effects which induce higher curvature corrections to the semiclassical gravity action on the brane. On the other hand, in the large tension limit, when the brane is close to the boundary of the bulk $AdS_4$, one recovers the usual BTZ black hole solution and the graviton becomes effectively massless.

The purpose of the present paper is to compute the holographic timelike entanglement entropy (tEE) \cite{Doi:2022iyj}-\cite{Guo:2025pru} in 3d massive gravity living in the brane, which allows for an exact quantum BTZ (qBTZ) solution that is conjectured to be dual to thermal states in a 2d defect CFT living at the intersection of the brane and the boundary of the bulk $AdS_4$. In the context of $AdS_3/CFT_2$, the timelike entanglement entropy (tEE) follows from the simple analytic continuation \cite{Doi:2023zaf} of the original Ryu-Takayanagi (RT) prescription \cite{Ryu:2006bv}-\cite{Ryu:2006ef} for spacelike intervals. The analytic continuation is correct in 3d, as this can be verified independently of $CFT_2$ calculations \cite{Doi:2023zaf}. Following these developments, numerous directions have been further explored. Some of these directions include- (i) phase transition in confining QFTs \cite{Afrasiar:2024lsi}, tEE in supersymmetric QFTs in ($0+1$)d \cite{Roychowdhury:2025ukl}, extremal surface and tEE for the dS space \cite{Narayan:2022afv}, tEE as a probe of RG flow \cite{Grieninger:2023knz}, tEE in the context of AdS/BCFT \cite{Chu:2023zah}, and the construction of the RT surface for tEE \cite{Heller:2024whi}.

The 2d defect CFT lives inside the boundary 3d CFT mentioned above. In summary, we compute the tEE in the brane, and this is different from the usual Ryu-Takayanagi (RT) prescription \cite{Ryu:2006bv}-\cite{Ryu:2006ef} to calculate the entanglement entropy for the full 3d boundary CFT. In the RT example, the extremal surface extends into the bulk $AdS_4$, with its end points on the KR brane. On the other hand, in the present computation, the extremal surface is within the 3d brane with its end points at the intersection of the brane and the bulk $AdS_4$. In other words, the calculations in this paper are performed from the perspective of an observer who is sitting on the KR brane and \textit{not} form the perspective of the bulk $AdS_4$.

In the present paper, we follow a recent proposal \cite{Nunez:2025gxq}-\cite{NR2} that bypasses the analytic continuation ($T \rightarrow i T$) and produces the correct results. Here, $T$ measures the size of the (timelike) entangling region which should be treated as a timelike separation (or interval with $T^2<0$) in the Lorentzian signature.  We introduce a parameter $\lambda(=\pm 1)$ that produces a Euclidean or Lorentzian result, depending on the choice of its sign. The Euclidean EE can be obtained for $\lambda=+1$. On the other hand, tEE corresponds to the setting $\lambda=-1$  \cite{Nunez:2025gxq}. 

The big picture here is to probe (quantum) black hole singularities by means of extremal surfaces and to find its imprints on tEE. This computation would reveal the central charge associated with the boundary 2d defect CFT and in particular the non-perturbative corrections that the central charge receives through backreaction/quantum effects on the brane. 

The remainder of the paper is organized as follows. We present our set-up in Section \ref{Sec2}. We reproduce the classical BTZ results in Section \ref{Sec3}. Finally, we apply our machinery to qBTZ in Section \ref{Sec4}, where we present a framework that can be used to compute tEE non-perturbatively in the backreaction parameter. Our analysis reveals an instability in the extremal surface beyond some critical backreaction, leading to a phase transition. We use tEE to define central charge associated with 2d CFT living on the boundary and explore the effects of quantum backreaction on it. Our analysis reveals a sharp fall-off in the central charge with an increasing backreaction that is rooted in the disintegration of the extremal surface leading to a phase transition. Finally, we draw our conclusion in Section \ref{Sec5}, where we outline some interesting future extensions of the present work.
\section{General algorithm}
\label{Sec2}
Quantum BTZ (qBTZ) solution was first constructed by authors in \cite{Emparan:2020znc}, where they consider an embedding of $AdS_3$ KR brane into bulk $AdS_4$ C metric \footnote{The $C$ metric \eqref{e1} describes an accelerating black hole in $AdS_4$ that intersects the brane orthogonally.}
\begin{align}
\label{e1}
	ds^2_{4}&=\frac{l^2}{(l+ x r)^2}\Big[-\mathcal{H}(r)dt^2+\frac{dr^2}{\mathcal{H}(r)}\nonumber\\&+r^2\Big(\frac{dx^2}{G(x)}+G(x)d\phi^2 \Big)\Big].
\end{align}

Here, we denote the above functions as
\begin{align}
	\mathcal{H}(r)=\kappa + \frac{r^2}{l^2_3}-\frac{\mu l}{r}, G(r)=1-\kappa x^2 - \mu x^3
\end{align}
where $\kappa = \pm 1, 0$ denotes different types of slicing of the brane \footnote{With $l=0$, different brane geometries are obtained for different choices of $\kappa$. For example, $\kappa=-1$ corresponds to classical BTZ solution and $\kappa=+1$ corresponds to global $AdS_3$ geometry on the brane.}, $l$ is the inverse acceleration ($A^{-1}$) of the black hole \footnote{When the brane is placed at $x=0$, its tension ($T_B$) goes inversely with $l$ \cite{Emparan:2020znc}.} and $\mu$ is the (positive) mass parameter. Moreover, here $l_3$ is the length scale related to the radius of curvature of $AdS_4$ as $L_4^{-2}=l^{-2}+l^{-2}_3$.

Putting the brane on $x=0$ hypersurface, the extrinsic curvature ($K_{ij}$) becomes proportional to the induced metric ($h_{ij}$) on the $AdS_3$KR brane and simplifies the embedding of the brane in bulk $AdS_4$ in the sense that Israel junction conditions \cite{israel} are trivially satisfied \cite{HosseiniMansoori:2024bfi}. 

This yields the qBTZ solution in the brane \cite{Emparan:2020znc}
\begin{align}
\label{e3}
	&ds^2_3=-f(r) dt^2 +\frac{dr^2}{f(r)}+r^2 d\phi^2 \\
	&f(r)=\frac{r^2}{l^2_3}-r^2_0-\frac{\nu l_3}{r} F(M)
	\label{e4}
	\end{align} 
where we identify the above constants as $r^2_0 =8 \mathcal{G}_3 M$ and $\nu = \frac{l}{l_3}$ is the measure of the quantum backreaction effect on the classical BTZ solution. Clearly, in the limit $\nu =0$, one recovers the classical BTZ solution \cite{HosseiniMansoori:2024bfi}. This is the limit in which the quantum backreaction becomes zero and the tension of the KR brane goes to infinity. On the other hand, at the opposite limit $\nu \gg 1 $, the backreaction becomes infinite and the tension of the brane goes to zero.

 Here, $ \mathcal{G}_3 $ is the renormalised Newton's constant in 3d and $M$ is the mass of the black hole. Finally, $ F(M) $ is another constant that can be expressed as a function of the smallest positive root ($ x_1 $) of $G(x)$. The above metric \eqref{e3} is an exact solution of the semiclassical Einstein equation (on the KR brane) to any order in the backreaction.

To begin with, we rewrite the metric \eqref{e3} following our latest convention \cite{Nunez:2025gxq}
\begin{align}
\label{e5}
ds^2_3= \lambda f(r) dt^2 +\frac{dr^2}{f(r)}+r^2 d\phi^2
\end{align}
where $\lambda=-1$ gives the metric in the Lorentzian signature. On the other hand, for $\lambda =+1$, we have the usual Euclidean metric with all positive signatures.

For our purpose, we repackage the function \eqref{e4} as
\begin{align}
\label{e6}
f(r)&=\frac{1}{r l^2_{3}}(r-r_+)(r^2+ r r_+ +\gamma)\\
r_+&=\sqrt{\gamma + r^2_0 l^2_{3}}
\end{align}
where $r_+$ is the location of the horizon and ($\bar\nu= \nu F(M)$)
\begin{align}
\label{e8}
	\gamma = \frac{\bar{\nu}l^3_3}{r_+}.
\end{align}
In our analysis, $\gamma$ is the measure of quantum backreaction in the KR brane. In what follows, we perform our analysis setting $l_3=1$, in which case $\gamma=\frac{\bar{\nu}}{r_+}$. Clearly, the classical solution corresponds to the setting $\gamma=0$.

Following \cite{Anegawa:2024kdj}, we set $\phi=$ constant and parametrise the geodesic as $t=t(\zeta)$ and $r=r(\zeta)$, which reveals the area functional of the present model as
\begin{align}
\label{e9}
\mathcal{A}=2\int_{\zeta_0}^{\zeta_b} d \zeta \sqrt{\lambda f(r)\dot{t}^2 +\frac{\dot{r}^2}{f(r)}}.
\end{align}

Here, $\zeta$ is an affine parameter such that $ \dot{t}=\frac{d t}{d \zeta} $ and $\dot{r}=\frac{dr}{d \zeta}$. Here, $ \zeta_0 $ is the affine parameter corresponding to the turning point at $r=r_c$ such that $ \dot{r}|_{\zeta =\zeta_0}=0 $. On the other hand, $ \zeta_b $ is the affine parameter corresponding to the boundary value $ r=r_b $. 

Clearly, there exists a (time) translation symmetry that yields a conserved charge ($E$), such that
\begin{align}
\label{e10}
f(r)\dot{t}=\frac{E}{\sqrt{\lambda}}
\end{align}
which is subject to the constraint \cite{Anegawa:2024kdj}
\begin{align}
\label{e11}
\lambda f(r)\dot{t}^2 +\frac{\dot{r}^2}{f(r)}=\lambda.
\end{align}

Using \eqref{e10} and \eqref{e11}, we find the following relation
\begin{align}
\label{e12}
\dot{r}^2+ E^2 = \lambda f(r).
\end{align}

The size of the entangling region can be obtained by integrating \eqref{e10}, which yields
\begin{align}
\label{e13}
T= \frac{2E}{\sqrt{\lambda}}\int_{r_c}^{r_b}\frac{dr}{\dot{r}f(r)}.
\end{align}

Clearly, in the Euclidean ($\lambda=+1$) signature
\begin{align}
    T_E = 2 E \int_{r_c}^{r_b}\frac{dr}{\dot{r}f(r)}.
\end{align}

On the other hand, in the Lorentzian ($\lambda=-1$) signature one finds
\begin{align}
    T_L = -2i E \int_{r_c}^{r_b}\frac{dr}{\dot{r}f(r)}=-i T_E
\end{align}
which is precisely an effect of the analytic continuation as proposed by the authors in \cite{Doi:2023zaf}, \cite{Anegawa:2024kdj}.

Finally, EE can be expressed as 
\begin{align}
\label{e16}
    S^{(\lambda)}_{EE}=\frac{\mathcal{A}^{(\lambda)}_{ext}}{4 \mathcal{G}_3}
\end{align}
where we denote the extremal surface as
\begin{align}
\label{e17}
    \mathcal{A}^{(\lambda)}_{ext}=2\sqrt{\lambda}\int_{r_c}^{r_b}\frac{dr}{\dot{r}}.
\end{align}
\section{Classical BTZ revisited}
\label{Sec3}
To begin with, we test our algorithm by revisiting the classical BTZ ($\gamma =0$) solution of \cite{Anegawa:2024kdj}. From \eqref{e12} we notice that $\dot{r}|_{r=r_c}=0$, which yields (in units $l_3=1$)
\begin{align}
\label{e18}
    r^2_c=r^2_+ + \lambda E^2
\end{align}
where $f(r)=r^2-r^2_+$.

The size \eqref{e13} of the entangling region is found to be
\begin{align}
\label{e19}
    T&=-\frac{2}{\sqrt{\lambda}r_+}\nonumber\\
    &\times \tan ^{-1}\left(\frac{-r \sqrt{\lambda  (r^2- r_c^2)}+\sqrt{\lambda } |r^2-r_+^2|}{E r_+}\right)\Big|^{r_b}_{r_c}.
\end{align}

One can approximate \eqref{e19} for a long extremal surface that extends from the boundary to the deep inside of the bulk such that $r_b \gg r_c \sim 0$, where $r_b \rightarrow \infty$. This yields
\begin{align}
\label{e20}
    T&=\frac{2}{\sqrt{\lambda}r_+}\tan ^{-1}\left(\frac{\sqrt{\lambda} r_+}{E}\right).
\end{align}

Notice that for $\lambda=+1$, this precisely reproduces the Euclidean result of \cite{Anegawa:2024kdj}, namely,
\begin{align}
\label{e21}
    T_E=\frac{2}{r_+}\tan ^{-1}\left(\frac{r_+}{E}\right).
\end{align}

On the other hand, in the Lorentzian ($\lambda=-1$) signature, one finds using \eqref{e20}
\begin{align}
\label{e22}
    T_L = \frac{2}{r_+}\tanh ^{-1}\left(\frac{r_+}{E}\right).
\end{align}

Next, we use the definition \eqref{e17} to compute the extremal surface, which yields 
\begin{align}
\label{e23}
 \mathcal{A}^{(\lambda)}_{ext}=-2  \log \left(\sqrt{\lambda  (r^2-  r_c^2)}-\sqrt{\lambda } r\right)\Big|^{r_b}_{r_c}.
\end{align}

Taking into account the long extremal surface as before, we notice from \eqref{e23}
\begin{align}
\label{e24}
    \mathcal{A}^{(\lambda)}_{ext}=2 \log \left(\frac{2r_b}{r_c}\right)=2 \log \left(\frac{2r_b}{\sqrt{r^2_+ + \lambda E^2}}\right).
\end{align}
which reproduces the Euclidean results of \cite{Anegawa:2024kdj} for $\lambda=+1$.

From \eqref{e21}, we find
\begin{align}
    \frac{r_+}{E}=\tan (\frac{r_+ T_E}{2})
\end{align}
which, when replaced with \eqref{e24}, yields
\begin{align}
    \mathcal{A}^{(\lambda=1)}_{ext}&= 2 \log \left(\frac{2r_b}{r_+\sqrt{1+  \frac{E^2}{r^2_+}}}\right)\nonumber\\
    &=2 \log \left(\frac{2r_b}{r_+}\sin(\frac{r_+ T_E}{2})\right).
\end{align}

To obtain tEE in the Lorentzian ($\lambda=-1$) signature, we invert \eqref{e22} to obtain
\begin{align}
\label{e27}
    \frac{r_+}{E}=\tanh (\frac{r_+ T_L}{2}).
\end{align}

Substituting \eqref{e27} into \eqref{e24}, we obtain
\begin{align}
\label{e28}
    \mathcal{A}^{(\lambda=-1)}_{ext}&= 2 \log \left(\frac{2r_b}{r_+\sqrt{1-  \frac{E^2}{r^2_+}}}\right)\nonumber\\
    &=2 \Big[\log \left(\frac{2r_b}{r_+}\sinh(\frac{r_+ T_L}{2})\right)+i\frac{\pi}{2}\Big].
\end{align}

Noting that the central charge \cite{Brown:1986nw}-\cite{Hartman:2008dq}
\begin{align}
\label{e29}
    \frac{c}{6}=\frac{1}{4 \mathcal{G}_3}
\end{align}
we finally obtain the tEE \eqref{e16}
\begin{align}
\label{e30}
    S^{(\lambda=-1)}_{EE}&=S_{tEE}\nonumber\\
    &=\frac{c}{3}\log \left(\frac{2r_b}{r_+}\sinh(\frac{r_+ T_L}{2})\right)+\frac{i\pi c}{6}.
\end{align}

In summary, the algorithm produces the correct EE and the central charge ($c$) in the Euclidean and Lorentzian signatures, previously found in \cite{Doi:2023zaf},\cite{Anegawa:2024kdj}. We use this machinery to explore qBTZ in the next section.
\section{Quantum BTZ and tEE}
\label{Sec4}
We carry out steps similar to classical BTZ example of the previous Section. Our purpose now would be to use \eqref{e12} for qBTZ and compute the extremal surface ($\mathcal{A}^{(\lambda)}_{ext}$) that leads to tEE in the Lorentzian ($\lambda=-1$) signature. 

The condition for the turning point \eqref{e12} yields \footnote{Notice, while solving the constraint \eqref{e31}, one encounters two complex and one real root. The real root signifies the existence of an extremal surface and has been considered in the computation. On the other hand, for a given backreaction ($\gamma$), one could probe deeper inside the black hole interior as the energy ($E$) of the configuration increases. In other words, the extremal surface with the larger energy ($E$) probes closer to the black hole singularity, which in turn sets $r_c \sim 0$.} (with $\lambda=-1$ in the Lorentzian signature)
\begin{align}
\label{e31}
    r_+=\frac{1}{2} \left(\sqrt{\frac{\left(\gamma +2 r_c^2\right)^2}{r_c^2}+4 E^2  }-\frac{\gamma }{r_c}\right)
\end{align}
which needs to be satisfied during computation. For a given horizon size ($r_+$) and the extremal surface with turning point ($r_c$), the corresponding energy ($E$) would be fixed from \eqref{e31}, where it will be decided by the (quantum) backreaction parameter ($\gamma$). We take this as input while evaluating the integrals \eqref{e13} and \eqref{e17}. Notice that in the classical ($\gamma=0$) limit, one recovers \eqref{e18} from \eqref{e31}.

Our staring points are the integrals \eqref{e13} and \eqref{e17} in the presence of quantum backreaction ($\gamma$), which in the Lorentzian ($\lambda =-1$) signature yield 
\begin{align}
\label{e32}
&\frac{|T_L|}{2E}= \int_{r_c}^{r_b}dr\frac{r\Big(\frac{  (r-r_+) (\gamma +r (r+r_+))}{r}+E^2 \Big)^{-1/2}}{(r-r_+) (\gamma +r (r+r_+))}\\
\label{e33}
&\mathcal{A}^{L}_{ext}=2\int_{r_c}^{r_b}\frac{dr}{\sqrt{\frac{(r-r_+) (\gamma +r (r+r_+))}{r}+E^2}}
\end{align}
where we denote $\mathcal{A}^{L}_{ext}=\mathcal{A}^{(\lambda=-1)}_{ext}$. Notice that $T_L$ in \eqref{e32} appears to have an overall -ve sign, and therefore we consider its magnitude only. Also, one should be able to check that for $\gamma = 0$, one precisely recovers \eqref{e22}. Furthermore, while evaluating the integral \eqref{e33}, the imaginary component is largely suppressed compared to the real component. As a consequence of this, we focus only on the real component of tEE (see Fig.\ref{fig1}(a)).

In the following, we add a Table \ref{table:1} to show the variation in the imaginary component of tEE ($\text{Im}S_{tEE}$) as a function of the backreaction parameter ($\gamma$). From the table below (Table \ref{table:1}), it can be clearly seen that the imaginary component of tEE ($\text{Im}S_{tEE}$) initially decreases with increasing backreaction. However, after a threshold has been reached, it starts to increase gradually.

\begin{table}[h!]
\centering
\begin{tabular}{||c c c c||} 
 \hline
 $r_b$ & $r_+$ & $\gamma$ & $\text{Im}S_{tEE}$ \\ [0.5ex] 
 \hline\hline
 10 & 0.2 & 0.001 & 7.44905 $\times 10^{-9} $ \\
 10 & 0.2 & 0.002 & 4.5849 $\times 10^{-9} $ \\
 10 & 0.2 & 0.003 & 3.24202 $\times 10^{-9} $ \\
 10 & 0.2 & 0.0035 & 3.74356 $\times 10^{-9} $\\
 10 & 0.2 & 0.004 & 4.5849 $\times 10^{-9} $\\
 10 & 0.2 & 0.005 & 6.48403 $\times 10^{-9} $\\[1ex] 
 \hline
\end{tabular}
\caption{Variation of the imaginary tEE with backreaction.}
\label{table:1}
\end{table}

\subsection{Stability and phase transition}
\label{pt}
Here, we pin down the aspect of stability of the extremal surface (against external perturbation \footnote{By external perturbation, here we always refer to the effects due to quantum backreaction on the KR brane.}) in a more rigorous way. This leads to a comparative study of the stability in the classical ($\gamma=0$) and quantum ($\gamma >0$) regime. We take the metric \eqref{e5} and define a codimension two hypersurface by choosing $t=t(r)$ and $\phi=0$. 

This yields the EE \eqref{e16} of the form \cite{Nunez:2025gxq}
\begin{align}
\label{ee34}
    S^{(\lambda)}_{EE}=\frac{1}{4\mathcal{G}_3}\int_{r_c}^{r_b}dr \sqrt{G^2(r)+F^2(r)t'^2(r)}
\end{align}
where we define $G(r)=\frac{1}{\sqrt{f(r)}}$ and $F(r)=\sqrt{\lambda}\sqrt{f(r)}$.

Following \cite{Faedo:2013ota}, we introduce the entity
\begin{align}
\label{ee35}
    Z(r_c)=\frac{d}{dr}\Big(\frac{\pi G(r)}{F'(r)} \Big)\Big|_{r=r_c}
\end{align}
where $Z(r_c)<0$ corresponds to a stable embedding. However, for $Z(r_c)>0$, the corresponding embedding is unstable against external perturbation \cite{Faedo:2013ota}.  

A straightforward computation reveals \footnote{In Lorentzian $\lambda=-1$ signature, one has to consider the imaginary component of \eqref{ee35}.}
\begin{align}
   Z(r_c)=\frac{4 \pi  r_c \left(\gamma  r_+ -r^3_c\right)}{\sqrt{\lambda } \left(2 r_c^3+\gamma  r_+\right)^2}.
\end{align}

Clearly, in the classical ($\gamma=0$) limit, $Z(r_c)=-\frac{\pi }{\sqrt{\lambda } r_c^2}<0$, which reflects the stability of the extremal surface. On the other hand, the backreaction in the brane induces instability of the configuration, and in particular, for a large backreaction, the entity \eqref{ee35} 
\begin{align}
    Z(r_c)\Big|_{\gamma\gg 1}=\frac{4 \pi  r_c}{\sqrt{\lambda } \gamma  r_+}+\mathcal{O}(\gamma^{-2})
\end{align}
is positive definite, thereby leading to instability. 

\begin{figure}
    \centering
    \includegraphics[width=1\linewidth]{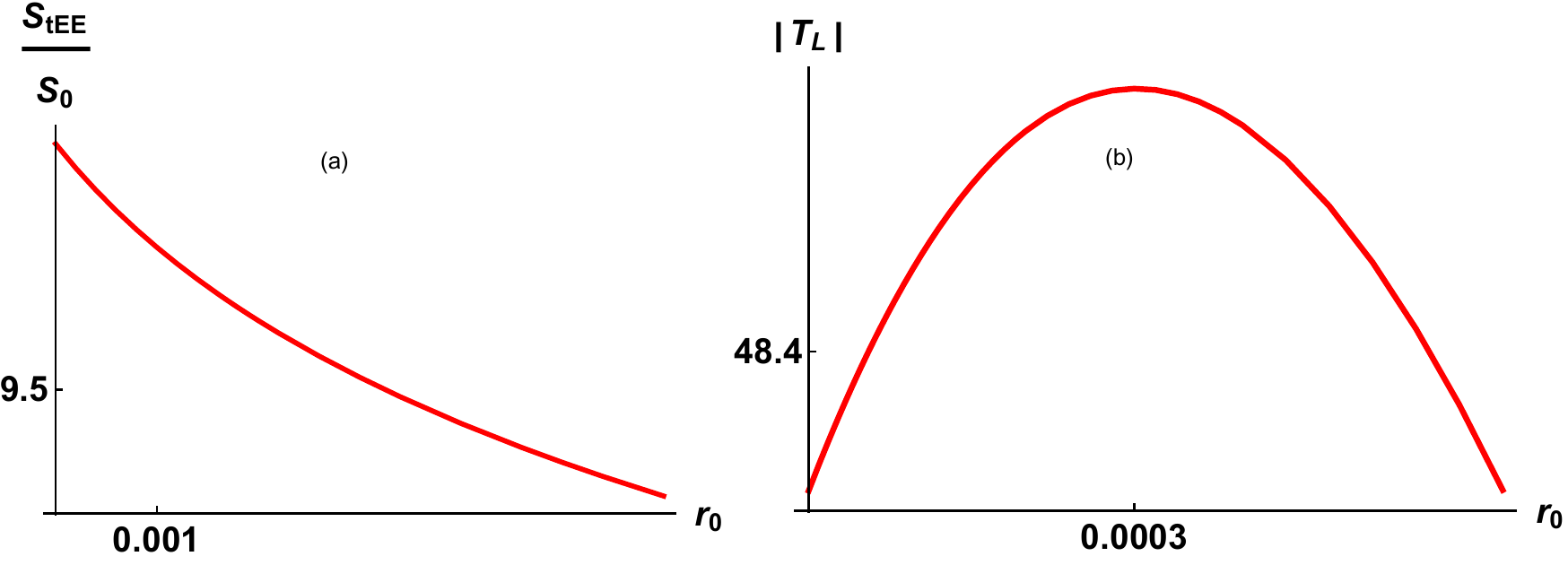}
    \caption{(a)Variation of tEE with backreaction parameter $r_0=\gamma r_+$, where we define $S_0=(4 \mathcal{G}_3)^{-1}$. (b) Variation of subsystem size with increasing backreaction.}
    \label{fig1}
\end{figure}

As we further elaborate, the change in sign in $Z(r_c)$ with increasing backreaction ($\gamma$) is related to the non-monotonicity of the interval \eqref{e32} (see Fig.\ref{fig1}(b)) and is indicative of an instability and, therefore, of a phase transition associated with the extremal surface \cite{Nunez:2025gxq}. The above idea is widely borrowed from previous studies on Wilson loops \cite{Faedo:2013ota}-\cite{Nunez:2009da}, which share remarkable similarities with \eqref{ee34}, at the level of action. However, we must emphasize that at this level this is purely an analogy in the sense that the origin of the phase transition in our example is quantum backreaction. On the other hand, in an example like \cite{Faedo:2013ota}, it is more geometric in the sense that the entity \eqref{ee35} changes sign depending on the value of the radial coordinate ($r_c$), which depends on the embedding.

In summary, we encounter a phase transition from a stable configuration of the extremal surface in the classical regime ($\gamma=0$) to an unstable configuration in the presence of quantum backreaction ($\gamma > 0$). The phase transition leads to a transition from the \textit{connected} extremal surface to a pair of \textit{disconnected} extremal surfaces, in the presence of a large quantum backreaction. 

\begin{figure}
    \centering
    \includegraphics[width=0.8\linewidth]{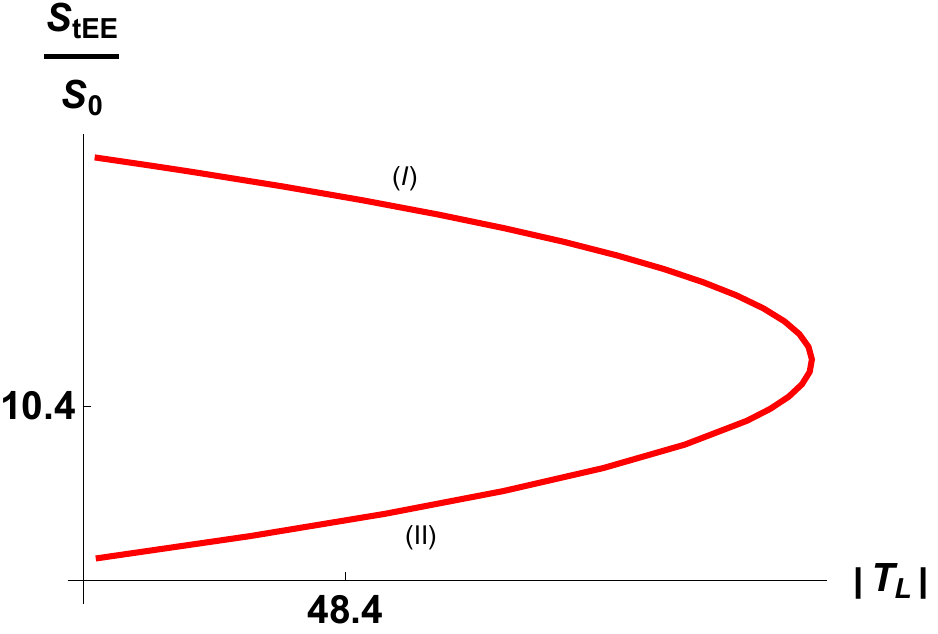}
    \caption{Timelike entanglement vs. subsystem size with increasing backreaction.}
    \label{fig2}
\end{figure}

To understand this more precisely, consider the backreaction to be small enough ($\gamma \ll 1$) such that one is infinitesimally away from the classical solution. In this regime, the transition can clearly be seen from a stable configuration of the extremal surface to an unstable configuration at a critical backreaction $\gamma_c=\frac{r^3_c}{r_+}$ (in units $l_3=1$). Clearly, for $\gamma<\gamma_c$, the configuration is stable, while it is unstable for $\gamma>\gamma_c$. In other words, for a given configuration of the extremal surfaces ($r_c$), there exists a point of instability at a critical backreaction $\gamma =\gamma_c$, beyond which the surface disintegrates into a pair of disconnected surfaces, leading to a phase transition.

We now move on to the exact estimation of the integrals \eqref{e32} and \eqref{e33}, where we set $r_+=0.2$, $r_b=10$ and $r_c=0.1$ in the following computation. We slowly turn on the backreaction parameter ($\gamma$) and obtain the corresponding energy (E) that satisfies the constraint \eqref{e31}. Clearly, for $\gamma>0.005$ the instability is triggered on and the extremal surface disintegrates into a pair of disconnected ones that has its end point on the hyperplane $r=r_c$. This corresponds to a critical value $r_0=\gamma r_+ =0.001$ in the parameter space.

In Fig.\ref{fig1}(a) and Fig.\ref{fig1}(b) we show the variation of tEE and the size of the subsystem with $r_0$ respectively, where the increase in the backreaction corresponds to an increase in $r_0$. Clearly, tEE decreases with an increase in backreaction, which is rooted in the instability of the extremal surface. On the other hand, the non-monotonicity of the size of the subsystem \eqref{e33} is indicative of instability and, therefore, a phase transition \cite{Klebanov:2007ws}-\cite{Kol:2014nqa}.

This is further clarified in Fig.\ref{fig2}, which reveals a double value in tEE \eqref{e33} for a given timelike separation \eqref{e32}. The upper arm (I) in Fig.\ref{fig2} corresponds to a stable configuration, and the corresponding backreaction is below the critical threshold ($\gamma < \gamma_c$). The lower arm (II) corresponds to a phase in which the instability is turned on due to an increasing backreaction on the KR brane. The instability would finally lead to a phase transition as the backreaction reaches the critical threshold ($\gamma \sim \gamma_c$). Notice that the critical threshold ($\gamma_c$) is different for different extremal surfaces, as it depends on the point of return $r_c$ in the bulk geometry. 

The phase transition is reminiscent of previously known results \cite{Klebanov:2007ws}-\cite{Kol:2014nqa} in the context of confining gauge theories and has recently been investigated for tEE in \cite{Nunez:2025gxq}. However, for the present gravity model, the dual QFT is not a confining one; instead it is a CFT$_2$ that (acts like a defect within CFT$_3$) lives at the interface of the brane and the bulk AdS$_4$. On top of that, the nature of the transition is different namely it is triggered due to quantum backreaction effects on the KR brane. However, in \cite{Klebanov:2007ws}-\cite{Kol:2014nqa} it is more geometric in nature, that is, the instability occurs beyond a certain critical value of the turning point ($r_c$) in the extremal surface.
\subsection{Central charge}
\label{cc}
We use tEE to define the central charge for the 2d CFT. It is evident that the central charge in higher derivative gravity should differ from the usual Einstein gravity \cite{Hu:2022ymx}. In a similar spirit, one should expect that the central charge of the 2d CFT, living at the intersection of the KR brane and the bulk $AdS_4$, must receive contributions from the higher derivative corrections in the bulk gravity action in 3d.

Given tEE, we define the central charge associated with the slab-like entangling region as \cite{Nunez:2025gxq}, \cite{Jokela:2025cyz}
\begin{align}
\label{e34}
    c_{slab}= T_L\partial_{T_L} S_{tEE}.
\end{align}

\begin{figure}
    \centering
    \includegraphics[width=0.8\linewidth]{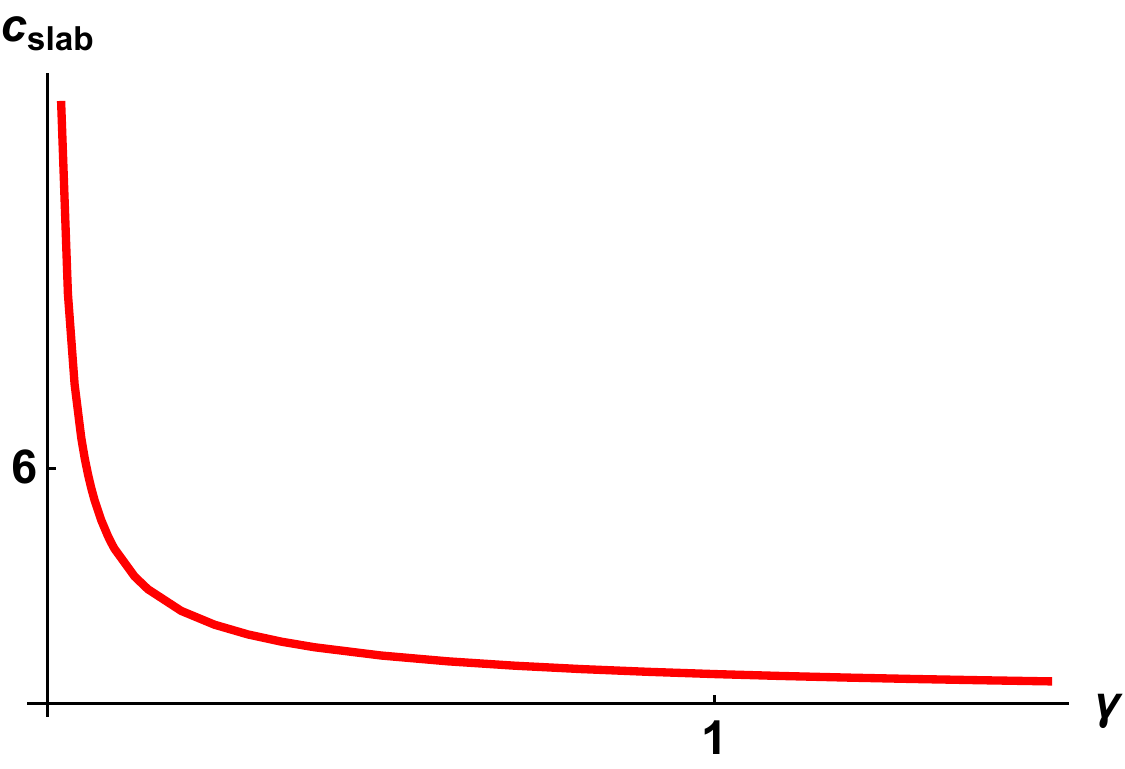}
    \caption{Variation of slab central charge with backreaction.}
    \label{fig4}
\end{figure}

Taking an example of the classical limit \eqref{e30}, this yields
\begin{align}
    c_{slab}=\kappa c
\end{align}
where $\kappa = \frac{r_+ E T_L}{12 r_b}\ll 1$ is the constant of proportionality. We can show that the combination $E T_L$ is independent of the energy of the configuration and is a pure number. For example, taking the small $T_L$ limit in \eqref{e27}, we find $E T_L=2$ and $\kappa=\frac{r_+}{6r_b}$, which depend only on the constant parameters ($r_+$ and $r_b$) of the problem.

The definition \eqref{e34} is in spirit quite similar to the Liu-Mezai central charge \cite{Jokela:2025cyz}, \cite{Liu:2012eea}. In Fig.\ref{fig4}, we show the variation of the central charge with the backreaction parameter ($\gamma$). As Fig.\ref{fig4} reveals, the central charge falls rapidly (from its classical value) when the backreaction starts increasing. Gradually, it reaches saturation once the backreaction is significant enough ($\gamma \sim 1$). The decrease in central charge \eqref{e34} is an artifact of the effects of quantum backreaction on the KR brane, which causes a phase transition of the extremal surface.

Like before, the analysis is performed for a given extremal surface, where we fix $r_c=0.1$, together with $r_+=0.2$ and $r_b=10$. The energy ($E$) of the configuration is obtained satisfying the constraint \eqref{e31}, where we vary the backreaction parameter ($\gamma$) independently. This is used as an input when computing the tEE using the extremal area \eqref{e33}. Considering that $S_{tEE}=S_{tEE}(r_c, E)$ and $T_L=T_L(r_c, E)$, the above procedure simply corresponds to the fact that the change in backreaction eventually changes the energy ($E$) of the configuration. Clearly, for some value of the backreaction parameter one exceeds the bound, namely $(\gamma r_+)^{1/3} >r_c(=0.1)$ and the extremal surface disintegrates due to the effects of backreaction causing a sharp fall in central charge \eqref{e34}. 

The derivative in \eqref{e34} is defined as the ratio of changes in the numerator to the denominator when measured with reference  to the classical BTZ solution, namely,
\begin{align}
    \frac{\Delta S_{tEE}}{\Delta T_L}=\frac{S_{tEE}|_{\gamma \neq 0}-S_{tEE}|_{\gamma =0}}{T_{L}|_{\gamma \neq 0}-T_{L}|_{\gamma =0}}.
\end{align}

Therefore, in summary, the calculation reveals that the central charge \eqref{e34} of the slab entangling region approaches the classical value \eqref{e29} for small backreaction ($\gamma \ll 1$), which is a large number in the limit $\mathcal{G}_3\ll 1$. Once the backreaction is turned on, it reduces this number that eventually saturates for $\gamma \geq 1$.
\section{Conclusion and outlook}
\label{Sec5}
In this paper, we build up the notion of holography for quantum BTZ (qBTZ) black holes by constructing an extremal surface and computing the entanglement entropy (EE) for the 2d CFT living at the boundary. We present an alternate recipe for computing the entanglement entropy for quantum BTZ (qBTZ) black holes in the Lorentzian signature. As we emphasize, our approach, in addition to producing the classical BTZ results on its own, provides an elegant way to address the computation of (timelike) entanglement entropy in the presence of quantum backreaction effects on the brane. 

Our analysis reveals that the timelike entanglement entropy (tEE) can be used as a probe of phase transition in the presence of quantum backreaction effects in a brane-world holographic setup. Beyond a certain critical backreaction ($\gamma> \gamma_c$), the extremal surface undergoes an instability that leads to phase transitions. This turns out to be quite similar in spirit as that of earlier observations in the context of confining gauge theories \cite{Klebanov:2007ws}-\cite{Kol:2014nqa}, with some of the major differences that are already alluded in the previous Section. It would be really nice to explore similar effects in a 2d CFT setup and, in particular, to explore whether quantum corrections can lead to instabilities, which should be observed through tEE. 

We also compute the central charge \eqref{e34} of the slab-like entangling region at the boundary, which decreases due to the quantum backreaction effects on the brane. As our analysis reveals, the fall-off in the central charge is an artifact of the disintegration of the extremal surface due to large backreaction effects on the brane. Before we conclude, here we outline some interesting future extensions of the present analysis. It would be nice to repeat the above calculations for charged \cite{Climent:2024nuj}-\cite{Feng:2024uia} and rotating qBTZ solutions \cite{Emparan:2020znc}-\cite{Bhattacharya:2025tdn}, particularly to explore the effects of rotation on tEE. This will reveal the central charge that is expected to receive additional corrections. Finally, it would be nice to understand the interpolation between the extremal surface that lives in the KR brane and the one that extends in the bulk $AdS_4$, hanging from the brane. These questions will be addressed in future work.
\section*{Acknowledgments} I thank Carlos Nunez and Juan Pedraza for their comments on the draft. DR acknowledges The Royal Society, UK for financial assistance. DR also acknowledges the Mathematical Research Impact Centric Support (MATRICS) grant (MTR/2023/000005) received from ANRF, India.



\begin{thebibliography}{99}
\bibitem{Emparan:2002px}
R.~Emparan, A.~Fabbri and N.~Kaloper,
``Quantum black holes as holograms in AdS brane worlds,''
JHEP \textbf{08}, 043 (2002)
doi:10.1088/1126-6708/2002/08/043
[arXiv:hep-th/0206155 [hep-th]].

\bibitem{Emparan:1999wa}
R.~Emparan, G.~T.~Horowitz and R.~C.~Myers,
``Exact description of black holes on branes,''
JHEP \textbf{01}, 007 (2000)
doi:10.1088/1126-6708/2000/01/007
[arXiv:hep-th/9911043 [hep-th]].

\bibitem{Emparan:1999fd}
R.~Emparan, G.~T.~Horowitz and R.~C.~Myers,
``Exact description of black holes on branes. 2. Comparison with BTZ black holes and black strings,''
JHEP \textbf{01}, 021 (2000)
doi:10.1088/1126-6708/2000/01/021
[arXiv:hep-th/9912135 [hep-th]].

\bibitem{Emparan:2020znc}
R.~Emparan, A.~M.~Frassino and B.~Way,
``Quantum BTZ black hole,''
JHEP \textbf{11}, 137 (2020)
doi:10.1007/JHEP11(2020)137
[arXiv:2007.15999 [hep-th]].

\bibitem{Bhattacharya:2025tdn}
D.~Bhattacharya, R.~A.~Hennigar, R.~B.~Mann and M.~Zhang,
``Charged rotating quantum black holes,''
[arXiv:2506.19941 [hep-th]].

\bibitem{Emparan:2022ijy}
R.~Emparan, J.~F.~Pedraza, A.~Svesko, M.~Toma\v{s}evi\'c and M.~R.~Visser,
``Black holes in dS$_{3}$,''
JHEP \textbf{11}, 073 (2022)
doi:10.1007/JHEP11(2022)073
[arXiv:2207.03302 [hep-th]].

\bibitem{Panella:2023lsi}
E.~Panella and A.~Svesko,
``Quantum Kerr-de Sitter black holes in three dimensions,''
JHEP \textbf{06}, 127 (2023)
doi:10.1007/JHEP06(2023)127
[arXiv:2303.08845 [hep-th]].

\bibitem{Climent:2024nuj}
A.~Climent, R.~Emparan and R.~A.~Hennigar,
``Chemical potential and charge in quantum black holes,''
JHEP \textbf{08}, 150 (2024)
doi:10.1007/JHEP08(2024)150
[arXiv:2404.15148 [hep-th]].

\bibitem{Feng:2024uia}
Y.~Feng, H.~Ma, R.~B.~Mann, Y.~Xue and M.~Zhang,
``Quantum charged black holes,''
JHEP \textbf{08}, 184 (2024)
doi:10.1007/JHEP08(2024)184
[arXiv:2404.07192 [hep-th]].

\bibitem{Panella:2024sor}
E.~Panella, J.~F.~Pedraza and A.~Svesko,
``Three-Dimensional Quantum Black Holes: A Primer,''
Universe \textbf{10}, no.9, 358 (2024)
doi:10.3390/universe10090358
[arXiv:2407.03410 [hep-th]].

\bibitem{Cartwright:2024iwc}
C.~Cartwright, U.~G\"ursoy, J.~F.~Pedraza and G.~Planella Planas,
``Perturbing a quantum black hole,''
JHEP \textbf{03}, 039 (2025)
doi:10.1007/JHEP03(2025)039
[arXiv:2408.08010 [hep-th]].

\bibitem{Frassino:2024bjg}
A.~M.~Frassino, R.~A.~Hennigar, J.~F.~Pedraza and A.~Svesko,
``Quantum Inequalities for Quantum Black Holes,''
Phys. Rev. Lett. \textbf{133}, no.18, 181501 (2024)
doi:10.1103/PhysRevLett.133.181501
[arXiv:2406.17860 [hep-th]].

\bibitem{Wu:2024txe}
S.~P.~Wu and S.~W.~Wei,
``Thermodynamical topology of quantum BTZ black hole,''
Phys. Rev. D \textbf{110}, no.2, 024054 (2024)
doi:10.1103/PhysRevD.110.024054
[arXiv:2403.14167 [gr-qc]].

\bibitem{Xu:2024iji}
Z.~M.~Xu, P.~P.~Zhang, B.~Wu and X.~Zhang,
``Thermodynamic bounce effect in quantum BTZ black hole,''
JHEP \textbf{12}, 181 (2024)
doi:10.1007/JHEP12(2024)181
[arXiv:2407.08241 [gr-qc]].

\bibitem{Frassino:2023wpc}
A.~M.~Frassino, J.~F.~Pedraza, A.~Svesko and M.~R.~Visser,
``Reentrant phase transitions of quantum black holes,''
Phys. Rev. D \textbf{109}, no.12, 124040 (2024)
doi:10.1103/PhysRevD.109.124040
[arXiv:2310.12220 [hep-th]].

\bibitem{Cartwright:2025fay}
C.~Cartwright, U.~G{\"u}rsoy, J.~F.~Pedraza and A.~Svesko,
``Quantum induced superradiance,''
[arXiv:2501.17231 [hep-th]].

\bibitem{Frassino:2025buh}
A.~M.~Frassino, R.~A.~Hennigar, J.~F.~Pedraza and A.~Svesko,
``Quantum censors: backreaction builds horizons,''
[arXiv:2505.09689 [hep-th]].

\bibitem{HosseiniMansoori:2024bfi}
S.~A.~Hosseini Mansoori, J.~F.~Pedraza and M.~Rafiee,
``Criticality and thermodynamic geometry of quantum BTZ black holes,''
Phys. Rev. D \textbf{111}, no.2, 024012 (2025)
doi:10.1103/PhysRevD.111.024012
[arXiv:2403.13063 [hep-th]].

\bibitem{Johnson:2023dtf}
C.~V.~Johnson and R.~Nazario,
``Specific heats for quantum BTZ black holes in extended thermodynamics,''
Phys. Rev. D \textbf{110}, no.10, 106004 (2024)
doi:10.1103/PhysRevD.110.106004
[arXiv:2310.12212 [hep-th]].

\bibitem{Nazario:2025qhl}
R.~Nazario,
``Specific Heats for Rotating Quantum BTZ Black Holes in Extended Thermodynamics,''
[arXiv:2502.02156 [hep-th]].

\bibitem{Kolanowski:2023hvh}
M.~Kolanowski and M.~Toma\v{s}evi\'c,
``Singularities in 2D and 3D quantum black holes,''
JHEP \textbf{12}, 102 (2023)
doi:10.1007/JHEP12(2023)102
[arXiv:2310.06014 [hep-th]].

\bibitem{Emparan:2021hyr}
R.~Emparan, A.~M.~Frassino, M.~Sasieta and M.~Toma\v{s}evi\'c,
``Holographic complexity of quantum black holes,''
JHEP \textbf{02}, 204 (2022)
doi:10.1007/JHEP02(2022)204
[arXiv:2112.04860 [hep-th]].

\bibitem{Chen:2023tpi}
B.~Chen, Y.~Liu and B.~Yu,
``Holographic complexity of rotating quantum black holes,''
JHEP \textbf{01}, 055 (2024)
doi:10.1007/JHEP01(2024)055
[arXiv:2307.15968 [hep-th]].

\bibitem{Steif:1993zv}
A.~R.~Steif,
``The Quantum stress tensor in the three-dimensional black hole,''
Phys. Rev. D \textbf{49}, 585-589 (1994)
doi:10.1103/PhysRevD.49.R585
[arXiv:gr-qc/9308032 [gr-qc]].

\bibitem{Shiraishi:1993qnr}
K.~Shiraishi and T.~Maki,
``Quantum fluctuation of stress tensor and black holes in three dimensions,''
Phys. Rev. D \textbf{49}, 5286-5294 (1994)
doi:10.1103/PhysRevD.49.5286
[arXiv:1804.07872 [gr-qc]].

\bibitem{Martinez:1996uv}
C.~Martinez and J.~Zanelli,
``Back reaction of a conformal field on a three-dimensional black hole,''
Phys. Rev. D \textbf{55}, 3642-3646 (1997)
doi:10.1103/PhysRevD.55.3642
[arXiv:gr-qc/9610050 [gr-qc]].

\bibitem{Casals:2019jfo}
M.~Casals, A.~Fabbri, C.~Mart\'\i{}nez and J.~Zanelli,
``Quantum-corrected rotating black holes and naked singularities in ( 2+1 ) dimensions,''
Phys. Rev. D \textbf{99}, no.10, 104023 (2019)
doi:10.1103/PhysRevD.99.104023
[arXiv:1902.01583 [hep-th]].

\bibitem{deHaro:2000wj}
S.~de Haro, K.~Skenderis and S.~N.~Solodukhin,
``Gravity in warped compactifications and the holographic stress tensor,''
Class. Quant. Grav. \textbf{18}, 3171-3180 (2001)
doi:10.1088/0264-9381/18/16/307
[arXiv:hep-th/0011230 [hep-th]].

\bibitem{Randall:1999ee}
L.~Randall and R.~Sundrum,
``A Large mass hierarchy from a small extra dimension,''
Phys. Rev. Lett. \textbf{83}, 3370-3373 (1999)
doi:10.1103/PhysRevLett.83.3370
[arXiv:hep-ph/9905221 [hep-ph]].

\bibitem{Randall:1999vf}
L.~Randall and R.~Sundrum,
``An Alternative to compactification,''
Phys. Rev. Lett. \textbf{83}, 4690-4693 (1999)
doi:10.1103/PhysRevLett.83.4690
[arXiv:hep-th/9906064 [hep-th]].

\bibitem{Karch:2000ct}
A.~Karch and L.~Randall,
``Locally localized gravity,''
JHEP \textbf{05}, 008 (2001)
doi:10.1088/1126-6708/2001/05/008
[arXiv:hep-th/0011156 [hep-th]].

\bibitem{Karch:2000gx}
A.~Karch and L.~Randall,
``Open and closed string interpretation of SUSY CFT's on branes with boundaries,''
JHEP \textbf{06}, 063 (2001)
doi:10.1088/1126-6708/2001/06/063
[arXiv:hep-th/0105132 [hep-th]].

\bibitem{Doi:2022iyj}
K.~Doi, J.~Harper, A.~Mollabashi, T.~Takayanagi and Y.~Taki,
``Pseudoentropy in dS/CFT and Timelike Entanglement Entropy,''
Phys. Rev. Lett. \textbf{130}, no.3, 031601 (2023)
doi:10.1103/PhysRevLett.130.031601
[arXiv:2210.09457 [hep-th]].

\bibitem{Doi:2023zaf}
K.~Doi, J.~Harper, A.~Mollabashi, T.~Takayanagi and Y.~Taki,
``Timelike entanglement entropy,''
JHEP \textbf{05}, 052 (2023)
doi:10.1007/JHEP05(2023)052
[arXiv:2302.11695 [hep-th]].

\bibitem{Nakata:2020luh}
Y.~Nakata, T.~Takayanagi, Y.~Taki, K.~Tamaoka and Z.~Wei,
``New holographic generalization of entanglement entropy,''
Phys. Rev. D \textbf{103}, no.2, 026005 (2021)
doi:10.1103/PhysRevD.103.026005
[arXiv:2005.13801 [hep-th]].

\bibitem{Li:2022tsv}
Z.~Li, Z.~Q.~Xiao and R.~Q.~Yang,
``On holographic timelike entanglement entropy,''
JHEP \textbf{04}, 004 (2023)
doi:10.1007/JHEP04(2023)004
[arXiv:2211.14883 [hep-th]].

\bibitem{Anegawa:2024kdj}
T.~Anegawa and K.~Tamaoka,
``Black hole singularity and timelike entanglement,''
JHEP \textbf{10}, 182 (2024)
doi:10.1007/JHEP10(2024)182
[arXiv:2406.10968 [hep-th]].

\bibitem{Guo:2025pru}
W.~z.~Guo and J.~Xu,
``A duality of Ryu-Takayanagi surfaces inside and outside the horizon,''
[arXiv:2502.16774 [hep-th]].

\bibitem{Ryu:2006bv}
S.~Ryu and T.~Takayanagi,
``Holographic derivation of entanglement entropy from AdS/CFT,''
Phys. Rev. Lett. \textbf{96}, 181602 (2006)
doi:10.1103/PhysRevLett.96.181602
[arXiv:hep-th/0603001 [hep-th]].

\bibitem{Ryu:2006ef}
S.~Ryu and T.~Takayanagi,
``Aspects of Holographic Entanglement Entropy,''
JHEP \textbf{08}, 045 (2006)
doi:10.1088/1126-6708/2006/08/045
[arXiv:hep-th/0605073 [hep-th]].

\bibitem{Afrasiar:2024lsi}
M.~Afrasiar, J.~K.~Basak and D.~Giataganas,
``Timelike entanglement entropy and phase transitions in non-conformal theories,''
JHEP \textbf{07}, 243 (2024)
doi:10.1007/JHEP07(2024)243
[arXiv:2404.01393 [hep-th]].

\bibitem{Roychowdhury:2025ukl}
D.~Roychowdhury,
``Holographic timelike entanglement and c theorem for supersymmetric QFTs in (0 + 1)d,''
JHEP \textbf{06}, 003 (2025)
doi:10.1007/JHEP06(2025)003
[arXiv:2502.10797 [hep-th]].

\bibitem{Narayan:2022afv}
K.~Narayan,
Phys. Rev. D \textbf{107}, no.12, 126004 (2023)
doi:10.1103/PhysRevD.107.126004
[arXiv:2210.12963 [hep-th]].

\bibitem{Grieninger:2023knz}
S.~Grieninger, K.~Ikeda and D.~E.~Kharzeev,
``Temporal entanglement entropy as a probe of renormalization group flow,''
JHEP \textbf{05}, 030 (2024)
doi:10.1007/JHEP05(2024)030
[arXiv:2312.08534 [hep-th]].

\bibitem{Chu:2023zah}
C.~S.~Chu and H.~Parihar,
``timelike entanglement entropy in AdS/BCFT,''
JHEP \textbf{06}, 173 (2023)
doi:10.1007/JHEP06(2023)173
[arXiv:2304.10907 [hep-th]].

\bibitem{Heller:2024whi}
M.~P.~Heller, F.~Ori and A.~Serantes,
``Geometric Interpretation of Timelike Entanglement Entropy,''
Phys. Rev. Lett. \textbf{134}, no.13, 131601 (2025)
doi:10.1103/PhysRevLett.134.131601
[arXiv:2408.15752 [hep-th]].

\bibitem{Nunez:2025gxq}
C.~Nunez and D.~Roychowdhury,
``Timelike Entanglement Entropy: a top-down approach,''
[arXiv:2505.20388 [hep-th]].

\bibitem{NR2}
C.~Nunez and D.~Roychowdhury,
``Interpolating between Space-like and Time-like Entanglement via Holography,''
[arXiv:2507.17805 [hep-th]].

\bibitem{israel}
W. Israel, Singular hypersurfaces and thin shells in general relativity, Nuovo Cim. B 44S10 (1966) 1.

\bibitem{Brown:1986nw}
J.~D.~Brown and M.~Henneaux,
``Central Charges in the Canonical Realization of Asymptotic Symmetries: An Example from Three-Dimensional Gravity,''
Commun. Math. Phys. \textbf{104}, 207-226 (1986)
doi:10.1007/BF01211590

\bibitem{Hartman:2008dq}
T.~Hartman and A.~Strominger,
``Central Charge for AdS(2) Quantum Gravity,''
JHEP \textbf{04}, 026 (2009)
doi:10.1088/1126-6708/2009/04/026
[arXiv:0803.3621 [hep-th]].

\bibitem{Faedo:2013ota}
A.~F.~Faedo, M.~Piai and D.~Schofield,
``On the stability of multiscale models of dynamical symmetry breaking from holography,''
Nucl. Phys. B \textbf{880}, 504-527 (2014)
doi:10.1016/j.nuclphysb.2014.01.016
[arXiv:1312.2793 [hep-th]].

\bibitem{Nunez:2009da}
C.~Nunez, M.~Piai and A.~Rago,
``Wilson Loops in string duals of Walking and Flavored Systems,''
Phys. Rev. D \textbf{81}, 086001 (2010)
doi:10.1103/PhysRevD.81.086001
[arXiv:0909.0748 [hep-th]].

\bibitem{Klebanov:2007ws}
I.~R.~Klebanov, D.~Kutasov and A.~Murugan,
``Entanglement as a probe of confinement,''
Nucl. Phys. B \textbf{796}, 274-293 (2008)
doi:10.1016/j.nuclphysb.2007.12.017
[arXiv:0709.2140 [hep-th]].

\bibitem{Kol:2014nqa}
U.~Kol, C.~Nunez, D.~Schofield, J.~Sonnenschein and M.~Warschawski,
``Confinement, Phase Transitions and non-Locality in the Entanglement Entropy,''
JHEP \textbf{06}, 005 (2014)
doi:10.1007/JHEP06(2014)005
[arXiv:1403.2721 [hep-th]].

\bibitem{Hu:2022ymx}
Q.~L.~Hu, D.~Li, R.~X.~Miao and Y.~Q.~Zeng,
``AdS/BCFT and Island for curvature-squared gravity,''
JHEP \textbf{09}, 037 (2022)
doi:10.1007/JHEP09(2022)037
[arXiv:2202.03304 [hep-th]].

\bibitem{Jokela:2025cyz}
N.~Jokela, J.~Kastikainen, C.~Nunez, J.~M.~Pen{\'\i}n, H.~Ruotsalainen and J.~G.~Subils,
``On entanglement c-functions in confining gauge field theories,''
[arXiv:2505.14397 [hep-th]].

\bibitem{Liu:2012eea}
H.~Liu and M.~Mezei,
``A Refinement of entanglement entropy and the number of degrees of freedom,''
JHEP \textbf{04}, 162 (2013)
doi:10.1007/JHEP04(2013)162
[arXiv:1202.2070 [hep-th]].

 \end{thebibliography}
\end{document}